\documentclass[showpacs,twocolumn,pra]{revtex4}
\usepackage{amsfonts}
\usepackage{amsmath}
\usepackage{amssymb}
\usepackage{graphicx}

\begin{document}

\title{Cavity-induced switching between localized and extended states in a
non-interacting Bose-Einstein condensate}
\author{Lu Zhou$^{1}$, Han Pu$^{2}$, Keye Zhang$^{1}$, Xing-Dong Zhao$^{3}$,
and Weiping Zhang$^{1}$}
\affiliation{$^{1}$Quantum Institute for Light and Atoms, Department of Physics, East
China Normal University, Shanghai 200062, China}
\affiliation{$^{2}$Department of Physics and Astronomy, and Rice Quantum Institute, Rice
University, Houston, TX 77251-1892, USA }
\affiliation{$^{3}$Department of Physics, Henan Normal University, Xinxiang 453007, China}

\begin{abstract}
We study an ultracold atom-cavity coupling system, which had been
implemented in experiment to display weak light nonlinearity [S. Gupta
\textit{et al}., Phys. Rev. Lett. \textbf{99}, 213601 (2007)]. The model is
described by a non-interacting Bose-Einstein condensate contained in a
Fabry-P\'{e}rot optical resonator, in which two incommensurate standing-wave
modes are excited and thus form a quasiperiodic optical lattice potential
for the atoms. Special emphasis are paid to the variation of atomic
wavefunction induced by the cavity light field. We show that bistability
between the atomic localized and extended states can be generated under
appropriate conditions.
\end{abstract}

\pacs{42.65.Pc, 42.50.Wk, 71.23.An, 72.15.Rn}
\maketitle

\section{introduction}

\begin{figure}[tbh]
\includegraphics[width=8cm]{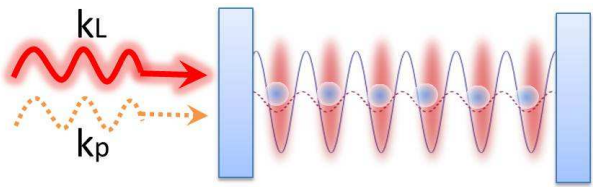}
\caption{{\protect\footnotesize (Color online) Schematic diagram showing the
system under consideration.}}
\label{scheme}
\end{figure}

We consider the following model depicted in Fig.~\ref{scheme}: A scalar
Bose-Einstein condensate (BEC) with atomic number $N$ is confined in a
high-finesse Fabry-P\'{e}rot cavity along the cavity axis in the $x$%
-direction, in which two standing-wave modes are excited. The atomic motion
in the transverse direction is restricted to the ground-state of a strong
harmonic trapping potential. At zero temperature the condensate can be
described by the wavefunction $\psi\left( x,t\right) $. The interaction
between the two cavity modes and atoms trapped inside is of dispersive
nature. One cavity mode is relatively strong and actively locked through
cavity feedback, thus providing a time-independent primary optical lattice
potential for the atoms. We call this mode as the trap mode, characterized
by the wavenumber $k_{L}$ and which is not affected by the atomic dynamics.
The second cavity mode with frequency $\omega_{c}$ and wavevector $%
k_{p}=\beta k_{L}$ is relatively weak and is influenced by the atomic
dynamics. Thus this mode can serve as a probe, which is driven by a coherent
laser field with frequency $\omega_{p}$ and amplitude $\eta$. We will
therefore call this mode as the probe mode. When $\beta$ is an irrational
number, the two cavity modes are incommensurate with each other, and
together they form a quasiperiodic lattice for the atoms.

Such a model has already been realized in experiment, with the emphasis on
the cavity electrodynamics \cite{stamper-kurn} and cavity optomechanics \cite%
{stamper-kurn1} (which assumes that the atomic deviation from its
equilibrium position is small). Due to the incommensurance of the trap mode
and probe mode, atoms in each well of the primary lattice is subject to a
different optical force induced by the probe mode. The collective effect in
turn induces a Kerr-type nonlinearity (proportional to the intensity of the
probe mode) to the probe-pump detuning. The strong nonlinearity thus allows
for optical bistability at photon numbers below unity.

On the other hand, it is well-known that the wavefunction describing
particle motion in a periodic potential with disorder can generally be
grouped into extended states and localized states. The transition between
them was originally studied by Anderson in the system of non-interacting
electrons in a crystal with impurities \cite{anderson localization}.
Anderson localization describes the localization of waves in disordered
media, which is considered as not only a fundamental phenomenon in
condensed-matter physics but also a universal phenomenon in many areas of
physics, such as light propagation in photonic crystal \cite{al in optics}.
The combination of ultracold atoms and optical potentials have provided a
powerful playground for the study of disorder-related phenomena \cite{review
article}. Specifically, in the case of an optical lattice a pseudorandom
potential can be generated by superimposing two standing waves of
incommensurate wavelengths. The localization of atomic matter wave in such a
quasiperiodic optical lattice have been experimentally observed \cite%
{roati08,deissler10} and extensively studied in theory \cite%
{damski03,modugno09,boers07,adhikari09,biddle10,roux08,cetoli10,fontanesi09}.

In the localized regime, atoms accumulate in one or several lattice sites,
thus strongly deviate from its equilibrium position. Some questions then
naturally arise: Whether the localization of the atomic wavefunction can be
observed in the atom-cavity coupling system described above? What is the
effect of cavity on localization? Will optical bistability still take place
in the localized regime?

Although the study of interplay between the atomic external or internal
degrees of freedom and the cavity light field have already attracted a lot
of theoretical \cite{self-organization,double
well,horak00,domokos02,moore99,larson08,meystre10,mekhov07,simons09,zhou09,pu11,larson11}
and experimental interest \cite{cavity with bec,esslinger09,esslinger08},
the new interest in this model lies in how the localization transition of
the atoms gets modified inside the optical resonator. The quasi-disorder
induced by the probe mode now depends on the atomic distribution, while the
properties of the atomic wavefunction is intimately related to disorder. The
problem is then highly nonlinear. Through this work, we expect to shed some
light on the problems related to the interplay between nonlinearity and
disorder.

The rest of the paper is organized as follows. Sec II introduces the
theoretical model, we show that the system can be mapped into the Aubry-Andr%
\'{e} model in the tight-binding limit. Sec III is devoted to the discussion
of a special case, in which the physical properties of the system can be
captured by a cavity-dressed asymmetric double-well system. Both the
equilibrium properties and dynamics are discussed. The more general case is
considered in Sec IV, in which we present numerical results and display the
bistability between the atomic extended and localized states.\ Finally we
conclude in Sec V.

\section{model}

In order to concentrate on the nonlinearity induced by the cavity light
field, we assume that the atomic s-wave scattering length is tuned to zero
by means of Feshbach resonance \cite{tunable interaction}. With the energy
measured in units of $\hbar\omega_{R}=\hbar^{2}k_{L}^{2}/2m$ $\left(
\omega_{R}\sim 2\pi\times5\text{ kHz for }^{87}\text{Rb atoms}\right) $\ and
length scaled in units of $k_{L}^{-1}$, the Gross-Pitaevskii (GP) equation
describing the evolution of the condensate wavefunction and the
corresponding mean-field equations of motion for the cavity light field read
\begin{subequations}
\label{em}
\begin{align}
i\frac{d}{dt}\psi\left( x,t\right) & =\left[ -\frac{d^{2}}{dx^{2}}%
+U_{t}\sin^{2}\left( x\right) \right.  \notag \\
& \left. +\left\vert \alpha\left( t\right) \right\vert ^{2}U_{0}\sin
^{2}\left( \beta x+\phi\right) \right] \psi\left( x,t\right) ,  \label{ema}
\\
\frac{d}{dt}\alpha\left( t\right) & =\left[ i\delta_{c}-iNU_{0}\left\langle
\sin^{2}\left( \beta x+\phi\right) \right\rangle -\kappa \right]
\alpha\left( t\right) +\eta.   \label{emb}
\end{align}
Here $\delta_{c}=\omega_{c}-\omega_{p}$ is the cavity-pump detuning, $%
U_{t\left( 0\right) }$ characterize the strength of atomic coupling with the
trap mode and probe mode, respectively. $\phi$ is the phase difference
between the two cavity modes. $\alpha$ represent the amplitude for the probe
mode with the decay rate $\kappa$. For the simplicity of the following
discussion, we assume $U_{t\left( 0\right) }>0$ and $\phi=0$. The bracket in
Eq.~(\ref{emb}) is defined as
\begin{equation*}
\langle f(x) \rangle= \int dx \,|\psi(x)|^{2} f(x) \,.
\end{equation*}


In the tight-binding limit, the atomic wavefunction $\psi\left( x\right) $
can be expanded over the lowest band Wannier basis $w_{j}\left( x\right)
=w\left( x-j \pi\right) $ defined by the primary lattice, i.e., $\psi\left(
x\right) =\sum_{j}c_{j}w_{j}\left( x\right) $ with $\sum_{j}\left\vert
c_{j}\right\vert ^{2}=1$. Retaining only the coupling between neighbouring
Wannier states and the onsite contribution of the probe mode, after
neglecting the constant terms, Eq.~(\ref{ema}) can be rewritten as
\end{subequations}
\begin{equation}
i\dot{c}_{j}=-J\left( c_{j+1}+c_{j-1}\right) -\Delta\cos\left( 2\pi\beta j
\right) c_{j},   \label{aa}
\end{equation}
with%
\begin{align*}
J & =-\int dx \, w_{j+1}\left( x\right) \left[ -\frac{d^{2}}{dx^{2}}%
+U_{t}\sin^{2}\left( x\right) \right] w_{j}\left( x\right) , \\
\Delta & =\frac{\left\vert \alpha\right\vert ^{2}U_{0}}{2}\int dx\cos\left(
2\beta x\right) \left\vert w\left( x\right) \right\vert ^{2}=\chi
U_{0}\left\vert \alpha\right\vert ^{2}.
\end{align*}
In estimating the value of $J$ and $\Delta$, we take the Gaussian
approximation for the wannier functions, which is valid for a sufficiently
deep primary lattice, and from which we have
\begin{subequations}
\begin{align}
J & = \frac{4}{\sqrt{\pi}} U_{t}^{3/4} \,\exp(-2\sqrt{U_{t}}) \,, \\
\Delta & = \frac{\left\vert \alpha\right\vert ^{2}U_{0}}{2} \,\exp(-\beta
^{2}/\sqrt{U_{t}}) \,.   \label{delta}
\end{align}

Taking the transformation
\end{subequations}
\begin{equation}
c_{j}=e^{ikj}\sum_{m}f_{m}e^{im\left( 2\pi\beta j \right) },
\label{transformation}
\end{equation}
Eq. (\ref{aa}) can be cast into the form%
\begin{equation}
i\dot{f}_{m}=-\frac{\Delta}{2}\left( f_{m+1}+f_{m-1}\right) -2J\cos\left(
2\pi\beta j+k\right) f_{m}.   \label{aa2}
\end{equation}
Equation (\ref{aa2}) has precisely the same form as Eq. (\ref{aa}) at $%
\Delta/J=2 $, which indicates the self-duality of the Aubry-Andr\'{e} (AA)
model \cite{aa}. Since Eq.~(\ref{transformation}) represents the typical
discrete Fourier transform which transforms localized states into extended
states and vice versa, $\Delta/J=2$ is thus regarded as the transition point
between the localized states and extended states.

The linear case without atomic feedback on the cavity mode is exactly the
same as those had been illustrated in \cite{modugno09}, one can observe the
AA localization by increasing the value of $\Delta/J$. The AA localization
resembles the Anderson localization of random systems and have been
experimentally observed with a non-interacting BEC in a bichromatic optical
lattice \cite{roati08}. In the following, we will investigate the nonlinear
effect on AA localization brought out by the atom-cavity interaction.

\section{superlattice}

\begin{figure}[tbh]
\includegraphics[width=8cm]{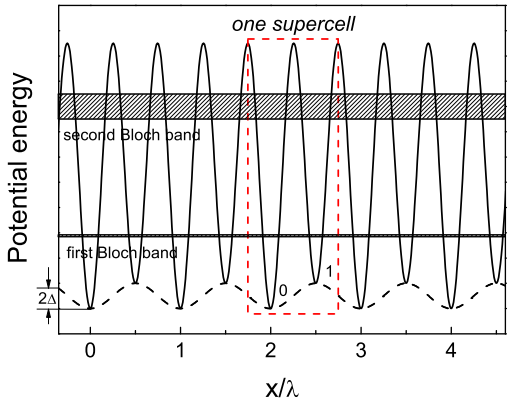}
\caption{{\protect\footnotesize The superlattice potential formed by the
cavity light field for }$\protect\beta =0.5${\protect\footnotesize , }$%
U_{t}=10${\protect\footnotesize \ and }$U_{0}\left\vert \protect\alpha%
\right\vert ^{2}=1${\protect\footnotesize . The stripes indicate the first
and second Bloch bands of the primary lattice formed by the trap mode. The
pertubation introduced by the probe mode is indicated by the dashed line.
The label 0(1) indicates the deep (shallow) site in a supercell.}}
\label{superlattice}
\end{figure}

Equation (\ref{aa}) indicates the periodicity $c_{j+1/\beta}=c_{j}$. In
order to gain some physical insights, we first consider the special case
that $1/\beta$ is an integer. Specifically, let us focus on the simple case
where $1/\beta=2$. The cavity modes then form a superlattice potential (more
specifically, a lattice of double-well potentials) for the condensed atoms,
as shown in Fig.~\ref{superlattice}. Such a potential has been realized in
the experiment \cite{nist}.

We have checked that, for the parameters considered here, the tight-binding
assumption is valid and all the population are in the lowest Bloch band.
Owning to the periodicity, the physics can be captured by one supercell in
the region of $x\in\left[ -\lambda/4,3\lambda/4\right] $, which represents a
cavity-mediated double-well system for the ultracold atoms. Similar systems
have also been discussed in Ref.~\cite{double well}. Due to the two
independent cavity mode used here, the double-well potential is \emph{%
asymmetric}, characterized by a deep well 0 and a shallow well 1.

\subsection{equilibrium properties: bistability between extended and
localized states}

Following Eq. (\ref{aa}), the effective equations of motion inside one
supercell can be written as%
\begin{align}
i\dot{c}_{0} & =-2Jc_{1}-\Delta c_{0},  \notag \\
i\dot{c}_{1} & =-2Jc_{0}+\Delta c_{1},   \label{reduced equation}
\end{align}
where we have used the periodic boundary condition $c_{j+2}=c_{j}$. This
equation illustrates a non-interacting BEC inside a double-well potential
with the tunneling coefficient $2J$ and the trap asymmetry $2\Delta$.\
Consider the following transformation according to Eq. (\ref{transformation})%
\begin{equation*}
c_{0}=f_{0}+f_{1}\text{, }c_{1}=f_{0}-f_{1},
\end{equation*}
the equations of motion for $f_{0}$ and $f_{1}$ reads%
\begin{align}
i\dot{f}_{0} & =-\Delta f_{1}-2Jf_{0},  \notag \\
i\dot{f}_{1} & =-\Delta f_{0}+2Jf_{1}.   \label{reduced equation1}
\end{align}
Eqs. (\ref{reduced equation1}) has precisely the same form as Eqs. (\ref%
{reduced equation}) with the role of $J$ and $\Delta$ interchanged, the two
equations are identical at $\Delta/J=2$, which signal the critical point.
The physical implication of the ciritical point can be captured via the
dimensionless parameters $z=2\int_{-\pi/2}^{\pi/2}dx\left\vert \psi\left(
x\right) \right\vert ^{2}-1\approx\left\vert c_{0}\right\vert
^{2}-\left\vert c_{1}\right\vert ^{2}$, which indicate the popultion
imbalance between the two wells. Under the initial condition of $c_{0}\left(
0\right) =1$, $c_{1}\left( 0\right) =0$ (i.e., with all atoms initially
prepared in the deep well), in the linear case where $J$ and $\Delta$ are
fixed external parameters, we have%
\begin{equation}
z\left( t\right) =1-\frac{2}{1+\left( \Delta/2J\right) ^{2}}\sin^{2}\sqrt{%
\Delta^{2}+4J^{2}}t.   \label{population imbalance}
\end{equation}
With the increase of $\Delta/2J$ from 0 to $\infty$, Eq. (\ref{population
imbalance}) shows that the minimum value of $z$ increases from $-1$ to 1,
indicating that the atomic wavefunction become localized and are less likely
to diffuse. At the critical value of $\Delta/J=2$, the value of $z$
oscillates between 0 and 1, which means that at most half of the atoms can
transport to the shallow well. In this sense we can group the atomic
wavefunction into the localized state and extended state.

In the nonlinear case, we will have to take the cavity feedback into
account. According to Eq.~(\ref{emb})%
\begin{equation}
\frac{d}{dt}\alpha\left( t\right) =\left[ i\Delta_{c}+iNU_{0}\chi z-\kappa%
\right] \alpha\left( t\right) +\eta,   \label{cavity}
\end{equation}
with $\Delta_{c}=\delta_{c}-NU_{0}/2$.\ By considering that the cavity decay
rate $\kappa$ is typically much larger than the atomic oscillation frequency
$J$, we adiabatically eliminate the probe field $\alpha$ from Eq.~(\ref%
{cavity}), replacing $\alpha$ in Eq.~(\ref{delta}) with%
\begin{equation}
\alpha=\frac{\eta}{\kappa-i\left[ \Delta_{c}+NU_{0}\chi z\right] }.
\label{cpn}
\end{equation}

\begin{figure}[tbh]
\caption{{\protect\footnotesize (Color online) The steady-state intensity of
the probe mode }$\left\vert \protect\alpha _{ss}\right\vert ^{2}$%
{\protect\footnotesize \ versus }$\protect\eta/\protect\kappa $%
{\protect\footnotesize \ using the two-mode treatment (red-dashed line) and
the self-consistent imagninary-time propagation (black dotted line). The
inset show }$\left\vert \protect\psi\left( x\right) \right\vert ^{2}$%
{\protect\footnotesize \ (solid line)\ versus the optical potential (dashed
line) inside one period for two different pump amplitude located at the two
sides of the bistable region. The vertical black dashed line indicate the
value of }$\left\vert \protect\alpha_{ss}\right\vert ^{2}$%
{\protect\footnotesize \ corresponding to }$\Delta/J=2$%
{\protect\footnotesize . Parameters: }$N=10^{4}${\protect\footnotesize , }$%
\protect\kappa=200\protect\omega_{R}${\protect\footnotesize , }$\protect%
\delta_{c}=0${\protect\footnotesize , }$U_{0}=0.2\protect\omega_{R}$%
{\protect\footnotesize . }$U_{t}=10\protect\omega_{R}${\protect\footnotesize %
.}} \label{bistability}\includegraphics[width=8cm]{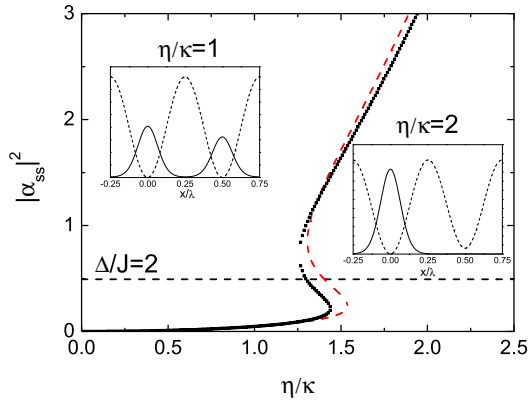}
\end{figure}

Equations (\ref{reduced equation}) can be rewritten in terms of $z$ and the
phase difference $\varphi$ between the amplitudes $c_{0}$ and $c_{1}$ as%
\begin{align}
\dot{z} & =4J\sqrt{1-z^{2}}\sin\varphi,  \notag \\
\dot{\varphi} & =-4J\frac{z}{\sqrt{1-z^{2}}}\cos\varphi+2\chi
U_{0}\left\vert \alpha\right\vert ^{2}.   \label{classical}
\end{align}
The steady-state solution of the system can then be solved according to the
above equations by takind the time derivative to be zero. An example is
shown by the red dashed line in Fig.~\ref{bistability}, which displays how
the mean photon number of the probe mode changes with the increase of
pumping amplitude $\eta/\kappa$. We can find three steady-state solutions in
certain parameter region, two of them are dynamically stable while the third
one is dynamically unstable, the unstable solution links the two stable
ones, representing a typical dispersive optical bistability.

The steady-state behavior can be understood as the interplay between the
double-well asymmetry and the cavity-induced nonlinearity, based on a
positive feedback mechanism: With the increase of the pumping amplitude $\eta
$ for the probe mode, its intensity increases as well. It results in a
larger $\Delta$ and thus enhance the double-well asymmetry felt by the
atoms. As a result, the atoms tend to stay in the deep well. This will give
rise to a larger population imbalance $z$ and shift the cavity resonance,
and will again lead to the increase of the probe mode intensity $\left\vert
\alpha\right\vert ^{2}$, as can be seen from Eq. (\ref{cpn}) by setting $%
\Delta_{c}<-NU_{0}\chi$.

In order to understand the variation of atomic wavefunction associated with
the optical bistability, we obtain the system steady-state in a
self-consistent manner by propagating Eq.~(\ref{ema}) in imaginary-time with
the steady-state value of the probe field amplitude from Eq.~(\ref{emb}).
The numerical simulation is performed on one period $\lambda$ with periodic
boundary condition. The result is displayed as the black dotted line in Fig.~%
\ref{bistability}, which shows good agreement with the two-mode result. The
typical wavefunctions are shown in the insets. In the linear mode without
cavity feedback, as the number of the probe photon increases, the atomic
system will experience a smooth crossover from the state that the atoms
evenly populated between the two wells to that only populated in the deep
well. Here in the cavity system, as the probe pumping rate $\eta/\kappa$ is
increased, the probe photon number will show discontinuous jumps.
Accompanying with this jump in photon number, the atomic wavefunction also
exhibits jumps between an extended state and a localized (self-trapped)
state.

\begin{figure}[tbh]
\includegraphics[width=8cm]{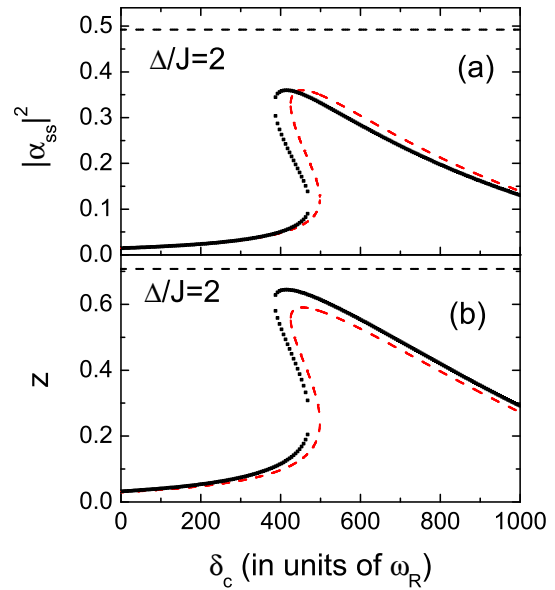}
\caption{{\protect\footnotesize (Color online) (a) The steady-state
intensity of the probe mode }$\left\vert \protect\alpha_{ss}\right\vert ^{2}$%
{\protect\footnotesize \ and (b) Inversion }$z${\protect\footnotesize \
versus cavity-pump detuning }$\protect\delta_{c}${\protect\footnotesize .
The red-dashed is the result from the two-mode treatment while the
black-dotted line results from the self-consistent imaginary time
propagation. The parameters are the same as Fig. \protect\ref{bistability}
except that }$\protect\eta=120\protect\omega_{R}${\protect\footnotesize .}}
\label{bistability2}
\end{figure}

We should point out here that the dispersive optical bistability is not
necessarily accompanied by the transition of the atomic wavefunction across
the critical point. An example is shown in Fig.~\ref{bistability2}. Here we
fix the pumping amplitude $\eta$ while varying the cavity-pump detuning $%
\delta_{c}$, as those had been done in experiment \cite{stamper-kurn}. It is
clear that the region for optical bistability is located below the critical
point, meaning that the bistability is taken place between extended states.
We anticipate that similar situation appears in the work of Ref.~\cite%
{stamper-kurn,stamper-kurn1}, in which the system operates with a weak probe
light field, the atomic deviation from its equilibrium position is small and
thus validates the mapping into cavity optomechanics.

\subsection{nonequilibrium dynamics}

\begin{figure}[tbh]
\includegraphics[width=9cm]{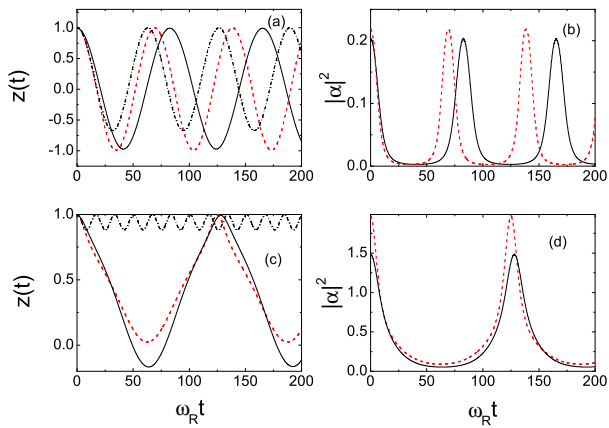}
\caption{{\protect\footnotesize (Color online) Evolution of population
imbalance }$z${\protect\footnotesize \ and intensity of the probe mode }$%
\left\vert \protect\alpha\right\vert ^{2}${\protect\footnotesize \ with time
}$t${\protect\footnotesize \ for }$\protect\eta/\protect\kappa=0.5$%
{\protect\footnotesize \ (upper panel) and }$1.5${\protect\footnotesize \
(lower panel). The other parameters are the same as those in Fig. \protect
\ref{bistability}. The black solid line are the results calculated with
numerical mithod (i) and red dashed is with method (ii). The dash dotted
line is given by Eq. (\protect\ref{population imbalance}).}}
\label{dynamics}
\end{figure}

In order to understand how the system dynamics are modified by the nonlinear
interaction, we use two different numerical methods: (i) integrate the GP
equation (\ref{ema}) in a self-consistent manner by assuming that the probe
field amplitude adiabatically follow the the variation of the condensate
wavefunction; (ii) use the two-mode model, and solve Eqs. (\ref{classical})
with Eq. (\ref{cpn}). To make a comparison with the dynamics predicted in
the linear case as represented by Eq.~(\ref{population imbalance}), we
assume that the atoms are initially localized in the deep well. The
numerical results are shown in Fig.~\ref{dynamics}.

Figs. \ref{dynamics}(a) and (b) are for $\eta/\kappa=0.5$, the other
parameters are the same as those in Fig. \ref{bistability}. In this case,
the probe field intensity is relatively low. As a result, the atomic
dynamics show little deviation from that described by Eq. (\ref{population
imbalance}). However when we increase $\eta/\kappa$ to the value of $1.5$,
the oscillation period of $z\left( t\right) $ experience a dramatic
enhancement compared to that predicted by Eq. (\ref{population imbalance}),
as can be seen from Fig. \ref{dynamics}(c). This critical slowing down takes
place when the system energy approaches a critical value of that defined by
an unstable saddle point, at which the contour line of energy changes its
topology from a closed to an open line. This behavior is usually associated
with bistability and has also been predicted in other systems before \cite%
{double well,zhou09}.

\begin{figure}[tbh]
\includegraphics[width=8cm]{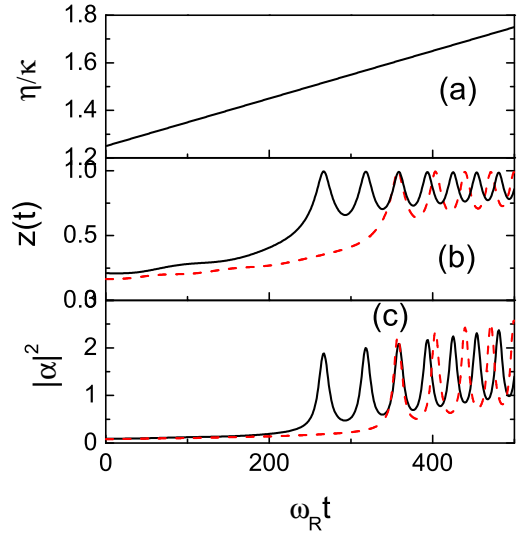}
\caption{{\protect\footnotesize (Color online) The pump amplitude (a),
population imbalance (b) and the probe intensity (c) versus time when the
pump amplitude is swept up. The parameters are the same as those in Fig.
\protect\ref{bistability}. The black solid line indicate the result got with
method (i) while the red dashed line use method (ii).}}
\label{transition-up}
\end{figure}

\begin{figure}[tbh]
\includegraphics[width=8cm]{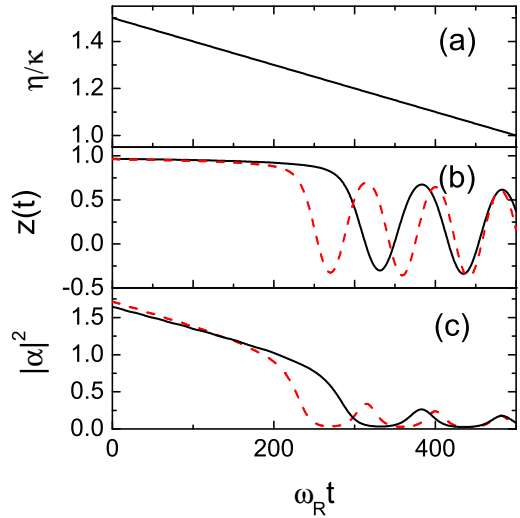}
\caption{{\protect\footnotesize (Color online) Same as Fig. \protect\ref%
{transition-up}, except that the pump amplitude is swept down.}}
\label{transition-down}
\end{figure}

It is self-evident that when the system is swept across the bistable region,
it cannot stay in the equilibrium solutions and non-steady state dynamics
will be excited. We implement this by ramping up the pump amplitude. The
population imbalance $z$ then experience a sharp transition to the localized
state and exhibit periodic oscillation with most atoms stay in the deep
well, as can be seen in Fig. \ref{transition-up}(b). If $\eta$ is swept in
the reverse direction, the value of $z$ can take the value of less than $0$,
which means that the system has entered into the extended state, as shown in
Fig. \ref{transition-down}(b). Comparing Figs. \ref{transition-up} and Figs. %
\ref{transition-down}, the abrupt transition exhibit a hysteretic behavior,
which indicates bistability and can be readily observed in experiment.

\section{quasiperiodic lattice: numerical results}

From the previous section we can understand that the interplay between
nonlinearity and disorder may lead to bistability between atomic extended
and localized states under appropriate conditions. In the more general case
that $1/\beta$ is not an integer, we can only resort to numerical simulation
to investigate the steady-state behavior of this system. Here we choose $%
\beta=\left( \sqrt{5}-1\right) /2$ as the inverse of the golden ratio. Then
the cavity modes form a quasiperiodic optical lattice potential for the
atoms.

\begin{figure}[tbh]
\includegraphics[width=8cm]{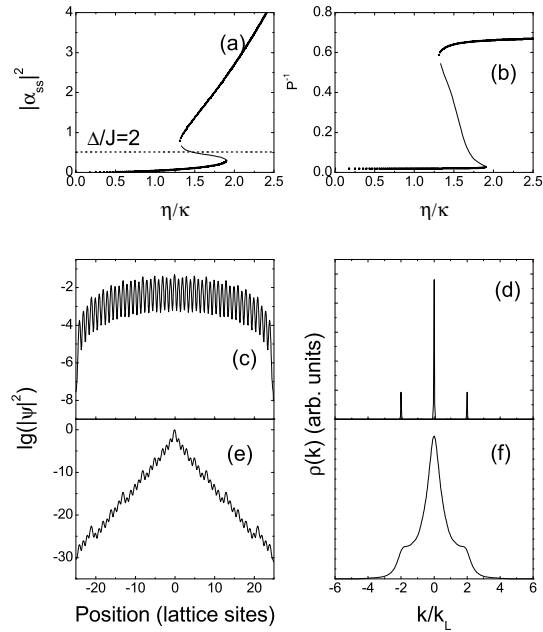}
\caption{{\protect\footnotesize Bistability between atomic extended and
localized states. (a) intracavity photon number for the probe mode }$%
\left\vert \protect\alpha_{ss}\right\vert ^{2}${\protect\footnotesize \ and
(b) inverse participation ratio }$P^{-1}${\protect\footnotesize \ plotted as
function of }$\protect\eta/\protect\kappa${\protect\footnotesize . Atomic
density (in log scale) and its momentum distribution for the lower bistable
branch (c)(d) and upper bistable branch (e)(f) at }$\protect\eta/\protect%
\kappa=1.5${\protect\footnotesize . The parameters are the same as those in
Fig. \protect\ref{bistability}.}}
\label{quasiperiodic}
\end{figure}

We focus first on the equilibrium properties of the system by numerically
solve Eq.~(\ref{em}) with 50 lattice sites. The numerical results are shown
in Fig.~\ref{quasiperiodic}. The intracavity photon number for the probe
mode as a function of the pumping rate is shown in Fig.~\ref{quasiperiodic}%
(a), which exhibits hysteretic behavior typical of dispersive optical
bistability. We characterize the property of atomic wavefunction with the
inverse participation ratio $P^{-1}=\int dx\left\vert \psi\left( x\right)
\right\vert ^{4}=\sum_{j}\left\vert c_{j}\right\vert ^{4}$, which is shown
in Fig.~\ref{quasiperiodic}(b). This quantity reflects the inverse of the
number of the lattice sites being occupied by the atoms. Hence a larger
value of $P_{-1}$ means the atoms are more localized in space. The atomic
density and momentum distribution of both the upper and lower bistable
branch at $\eta/\kappa=1.5$ are shown in Figs.~\ref{quasiperiodic}(c)-(f).
It is clear that in Figs.~\ref{quasiperiodic}(c) and (d) the atomic
wavefunction in the lower bistable branch display typical properties of
bloch extended states with almost uniform distribution, while the associated
momentum distribution $\rho\left( k\right) $ has well-defined peaks located
at $k=0$, $\pm2k_{L} $. The atomic density profile in the upper bistable
branch is exponentially localized, with the corresponding momentum
distribution $\rho\left( k\right) $ broadened, as can be seen in Figs. \ref%
{quasiperiodic}(e) and (f). The atomic momentum distribution in a cavity can
be readily observed via absorption imaging after the time-of-flight \cite%
{esslinger09}. This can serve as a clear evidence that the bistability can
really taken place between the atomic Bloch extended and AA localized states.

\begin{figure}[tbh]
\includegraphics[width=8cm]{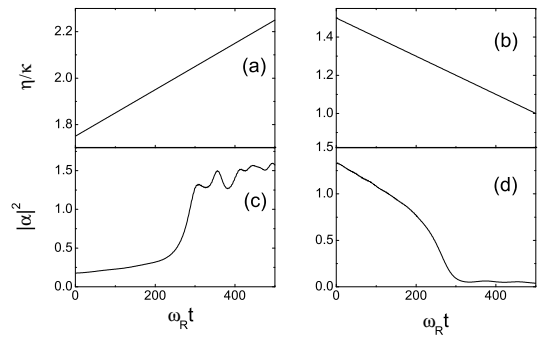}
\caption{{\protect\footnotesize The pump amplitude and the probe intensity
versus time when the pump amplitude is swept up (left column) and down
(right column). The parameters are the same as those used in Fig. \protect
\ref{quasiperiodic}.}}
\label{bistability3}
\end{figure}

We also investigate the transition dynamics from the extended states to
localized states and vice versa by ramping the pump amplitude up or down,
the results are shown in Fig.~\ref{bistability3}. Similar to the dynamics
displayed in Figs.~\ref{transition-up} and \ref{transition-down}, when the
pumping strength is ramped up to exceed a critical value, or ramped down to
below a lower threshold value, the system will become dynamically unstable
and the probe intensity is suddenly increased or decreased, typical of
dispersive optical bistability.

\begin{figure}[tbh]
\includegraphics[width=8cm]{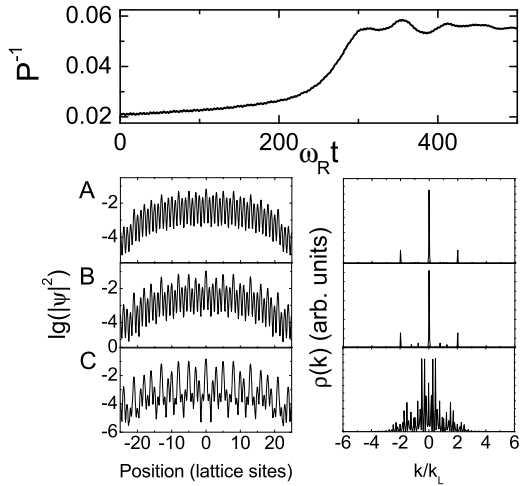}
\caption{{\protect\footnotesize Evolution of the inverse participation ratio
}$P^{-1}${\protect\footnotesize \ when the pump amplitude is swept up (the
Uppermost figure). The subsequent figure show the atomic density (in log
scale) (left column) and its momentum distribution (right column). The row
marked with A, B and C correspond to }$\protect\omega _{R}t=100,250$%
{\protect\footnotesize \ and }$350${\protect\footnotesize , respectively.}}
\label{e-l transition}
\end{figure}

\begin{figure}[tbh]
\includegraphics[width=8cm]{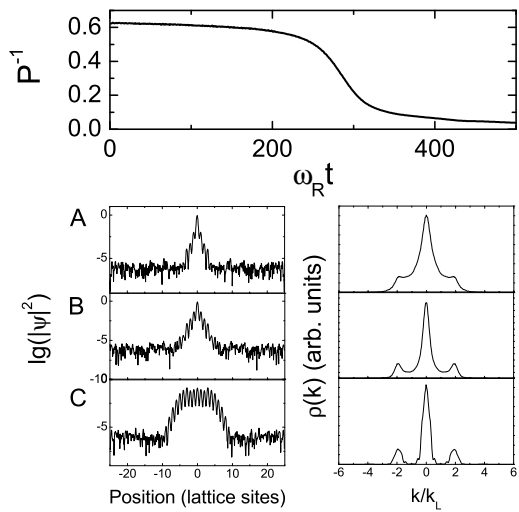}
\caption{{\protect\footnotesize Same as Fig. \protect\ref{e-l transition},
except that the pump amplitude is swept down. Row A, B and C correspond to }$%
\protect\omega_{R}t=100${\protect\footnotesize , }$250$%
{\protect\footnotesize \ and }$400${\protect\footnotesize , respectively.}}
\label{l-e transition}
\end{figure}

The corresponding atomic dynamics are shown in Figs.~\ref{e-l transition}
and \ref{l-e transition}. Fig. \ref{e-l transition} is for the case that the
pumping amplitude is ramped up. We start from an extended wavefunction,
whose momentum distribution has peaks located at $k=0$, $\pm2k_{L}$, as
shown in the row marked with A. Along with the increase of the pump
amplitude, the probe intensity gradually increases, the atomic spatial
distribution becomes more inhomogeneous, and a small fraction of atoms will
be scattered into momentum states $p=\pm2\hbar k_{p},\pm2\hbar\left(
k_{p}-k_{L}\right) ,\cdot \cdot\cdot$ due to the beating of the two
standing-wave cavity mode, as shown in the B row. When the pumping amplitude
becomes so strong that the probe intensity greatly increases and the atomic
BEC break into fragments, this is an intermediate atomic state between
extended states and localized states, as shown in row C. In this process,
the inverse participation ratio $P^{-1}$ finally jumped to a value around $%
0.05$, which is much lower than that would be expected for the present
system in the localized regime (around $0.6$ as indicated in Fig.~\ref%
{quasiperiodic}(b)). Comparing with the steady-state inverse participation
ratio as shown in Fig.~\ref{quasiperiodic}(b), one can see that the system
can follow the instantaneous steady state (the lower branch) initially, but
fails to do so after the pumping rate reaches the turning point of the lower
branch. Beyond this point, the probe mode intensity jumps up and the atomic
wavefunction, instead of forming a localized wavepacket as in the case of
the steady state, becomes fragmented.

By contrast, when the pumping amplitude is ramped down, the inverse
participation ratio $P^{-1}$ gradually decrease to a value approaching zero,
as shown in Fig.~\ref{l-e transition}. Contrary to Fig.~\ref{e-l transition}%
, it displays the behavior of approximately following the instantaneous
steady state adiabatically despite the occurance of bistable transition.
This is because that the atoms is initially in a localized state, whose
overlap with the probe mode is small, as can be seen from row A. Reducing
the probe intensity only lead to the weakening of disorder, the atomic
wavefunction diffuses and the sites neighbouring to the localized sites
start to be macroscopically occupied, as indicated in row B and C, and the
atoms becomes more and more delocalized, roughly following the steady
instantaneous steady state.

\section{conclusion}

In summary, we have investigated the system of a non-interacting BEC
dispersively coupled to two standing-wave cavity modes. The equilibrium
properties of the system are studied under the mean-field approximation. In
the special case of $1/\beta=2$, the cavity modes form a lattice of
asymmetric double-well which can be well described by a two-mode model. The
results obtained from the two-mode model and the self-consistent
imaginary-time propagation of the GP equation show very good agreement. Due
to the interplay between the nonlinearity and disorder, bistability between
the atomic extended and localized states can take place, which is associated
with the occurance of optical bistability. This is also confirmed via the
numerical simulation with an irrational $1/\beta$ in which case the cavity
modes form a quasiperiodic lattice.

\begin{acknowledgments}
We gratefully thank Prof. Hong Y. Ling for helpful discussions. This work
was supported by the National Basic Research Program of China (973 Program)
under Grant No. 2011CB921604, the National Natural Science Foundation of
China under Grant No. 11004057, No. 10828408 and No. 10588402, Shanghai
Leading Academic Discipline Project under Grant No. B480, the `Chen Guang'
project supported by Shanghai Municipal Education Commission and Shanghai
Education Development Foundation under Grant No. 10CG24, and the Fundamental
Research Funds for the Central Universities. HP is supported by the US NSF,
the Welch Foundation (Grant No. C-1669) and the DARPA OLE program.
\end{acknowledgments}

\end{document}